\documentclass[aps,physrev,reprint,superscriptaddress,floatfix,bibnotes]{revtex4-2}

\usepackage{graphicx}
\usepackage{dcolumn}
\usepackage{bm}

\usepackage[utf8]{inputenc}
\usepackage[T1]{fontenc}
\usepackage{graphicx}
\usepackage{dcolumn}
\usepackage{bm}
\usepackage[hidelinks]{hyperref}
\usepackage{amssymb}
\usepackage{amsmath}
\usepackage[figure]{hypcap}
\usepackage[version=4]{mhchem}
\usepackage{subscript}
\usepackage{siunitx}
\DeclareSIUnit\Ions{ions}
\DeclareSIUnit\electronvolt{eV}
\DeclareSIUnit\electrons{e^{-}}
\DeclareSIUnit\grooves{grooves}
\usepackage{placeins}
\usepackage{upgreek}
\usepackage{multirow}
\usepackage[modules]{chemmacros}
\chemsetup{modules = {reactions,isotopes} , greek   = upgreek ,formula = chemformula} 
\DeclareSIUnit\u{u}

\makeatletter
\def\@email#1#2{%
 \endgroup
 \patchcmd{\titleblock@produce}
  {\frontmatter@RRAPformat}
  {\frontmatter@RRAPformat{\produce@RRAP{*#1\href{mailto:#2}{#2}}}\frontmatter@RRAPformat}
  {}{}
}%
\makeatother
\begin{document}

\preprint{AIP/123-QED}

\title{\textbf{Multiply charged uranium monoxide as a versatile probe of fundamental physics}}%

\author{Jonas Stricker}
\email{Theory contact author: gaulkons@uni-mainz.de}
\affiliation{Department of Chemistry - TRIGA Site, Johannes Gutenberg University Mainz, 55099 Mainz, Germany}
\affiliation{PRISMA$^+$ Cluster of Excellence, Johannes Gutenberg University Mainz, 55099 Mainz, Germany}
\affiliation{Helmholtz Institute Mainz, 55099 Mainz, Germany}
\thanks{}

\author{Konstantin Gaul}
\email{Contact author: jostrick@uni-mainz.de}
\affiliation{Helmholtz Institute Mainz, 55099 Mainz, Germany}
\affiliation{Institut für Physik, Johannes Gutenberg University Mainz, 55099 Mainz, Germany}

\author{Paul Fischer}
\affiliation{Institut für Physik, Universität Greifswald, 17487 Greifswald, Germany}

\author{\\Lennard M. Arndt}
\affiliation{Department of Chemistry - TRIGA Site, Johannes Gutenberg University Mainz, 55099 Mainz, Germany}

\author{Florian Kraus}
\affiliation{Institut für Anorganische Chemie,
Universität Bonn, 53121 Bonn, Germany}

\author{David Krug}
\affiliation{Fachbereich Chemie,
Philipps-Universität Marburg, 35032 Marburg, Germany}

\author{Dennis Renisch}
\affiliation{Department of Chemistry - TRIGA Site, Johannes Gutenberg University Mainz, 55099 Mainz, Germany}
\affiliation{Helmholtz Institute Mainz, 55099 Mainz, Germany}

\author{\\Ferdinand Schmidt-Kaler}
\affiliation{PRISMA$^+$ Cluster of Excellence, Johannes Gutenberg University Mainz, 55099 Mainz, Germany}
\affiliation{Helmholtz Institute Mainz, 55099 Mainz, Germany}
\affiliation{QUANTUM, Johannes Gutenberg University Mainz, 55099 Mainz, Germany}

\author{Lutz Schweikhard}
\affiliation{Institut für Physik, Universität Greifswald, 17487 Greifswald, Germany}

\author{Jean Velten}
\affiliation{Department of Chemistry - TRIGA Site, Johannes Gutenberg University Mainz, 55099 Mainz, Germany}

\author{Christoph E. Düllmann}
\affiliation{Department of Chemistry - TRIGA Site, Johannes Gutenberg University Mainz, 55099 Mainz, Germany}
\affiliation{PRISMA$^+$ Cluster of Excellence, Johannes Gutenberg University Mainz, 55099 Mainz, Germany}
\affiliation{Helmholtz Institute Mainz, 55099 Mainz, Germany}
\affiliation{GSI Helmholtzzentrum für Schwerionenforschung GmbH, 64291 Darmstadt, Germany}

\date{\today}

\begin{abstract}

Multiply charged actinide molecules provide a unique platform to study fundamental physics and chemistry phenomena including fundamental symmetries and the nature of the chemical bond under extreme conditions. 
Inherently large relativistic effects associated with a high atomic number $Z$, can be further increased with increasing molecular charge. Such large relativistic effects strongly enhance the electronic sensitivity to symmetry-violating nuclear effects, including nuclear Schiff moments.
Experimental investigations of multiply charged actinide molecules are challenging because high charges destabilize chemical bonds promoting Coulomb explosion.
We demonstrate a method to systematically generate and detect molecular ions at the edge of chemical stability.
By applying high-fluence laser ablation to a depleted uranium metal foil, we produce atomic uranium ions \ce{U^{z+}} and uranium monoxide cations \ce{UO^{z+}} ($z = 1$ - 4).
We observe \ce{UO^{3+}}, which exhibits a comparatively simple electronic structure and is therefore promising for precision spectroscopy. The observation of \ce{UO^{4+}} allows to progress into a largely unexplored regime of actinide chemistry with metastable molecules.
Our experimental results are completed by fully-relativistic density functional theory calculations of equilibrium bond lengths, charge distributions, and binding energies of all observed molecules.
Calculations of symmetry-violating properties suggest a pronounced sensitivity of \ce{UO^{3+}} to hadronic $CP$ violation.
Our new approach opens a pathway for high-precision investigations of fundamental symmetries and the exploration of actinide chemistry in previously inaccessible regimes.

\end{abstract}

\maketitle

\section{Introduction}
The observed non-conservation of the combined symmetry of charge and spatial parity ($CP$) is considered to be insufficient for explaining the imbalance of matter and anti-matter in the universe (baryon asymmetry) \cite{sakharov:1967,canetti:2012}. Therefore, it is expected that $CP$-violation beyond the predictions of the Standard Model of particle physics exists \cite{chupp:2019}. Molecular precision spectroscopy is among the most sensitive probes of $CP$-violation \cite{demille:2015,roussy:2023}. As $CP$-violation sensitivity steeply increases with the nuclear charge number $Z$, i.e., the number of protons inside the atomic nucleus \cite{khriplovich:1997,safronova:2018}, actinide molecules are particular promising for this purpose \cite{andreev:2018}. Moreover, actinides often exhibit nuclear deformations which can further enhance $CP$-violating effects in the nucleus \cite{auerbach:1996}. In recent years, possibilities for precision spectroscopy of short-lived radioactive molecules increased interest in actinide molecules \cite{GarciaRuiz2020,ArrowsmithKron2024}. Most considerations of molecular spectroscopy were restricted to the early actinides Ac \cite{gaul:2019,skripnikov:2020,chen:2024,athanasakis-kaklamanakis:2025, Au2024} and Th \cite{fleig:2014,Skripnikov2013,Skripnikov2015,Skripnikov2015a,andreev:2018,gaul:2019, Ng2022} because of their simpler electronic structure compared to later actinides such as Pa, U, Np or Pu. The theoretical work has demonstrated that actinide fluoride ions (AnF$_z^{n+}$) with higher charge states $n \geq 2$ can exhibit large enhancement factors for precision measurements \cite{zulch:2022}, although their experimental stability and the implementation of suitable measurement schemes remain open. For example \ce{ThF^2+} \cite{Stricker2025} and \ce{PaF^3+} \cite{zulch:2022}  are isoelectronic to \ce{RaF}, which has a simple electronic structure ideally suited for precision experiments \cite{isaev:2010, Isaev2013, GarciaRuiz2020, Udrescu2021, udrescu:2024,wilkins:2025}. 

Oxide molecules can be directly produced from standard metal foils in large amounts without the requirements of either a dedicated synthesis or gas phase formation. However, the chemical binding in molecular oxide ions is expected to be weaker because the ionization energy of oxygen is considerably lower than that of fluorine \((\Delta E_\mathrm{IP}(\mathrm{O}) < \Delta E_\mathrm{IP}(\mathrm{F}))\) \cite{Schröder1999}. While actinide monoxides and dioxides have been studied in the gas phase, their accessible charge states have remained limited. 
Cornehl et al. demonstrated the formation of uranyl dications \ce{UO2^{2+}} from gas phase oxidation of uranium cations \cite{Cornehl1996}. This approach was extended across the actinide series, producing actinide monoxide dications  \ce{AnO^{2+}} (An = Th, U - Am) \cite{Jackson2002, gibson:2005} and actinide dioxide dications \ce{AnO2^{2+}} (An =Th, U - Pu) in ion traps \cite{gibson:2005}. Similar methods yielded doubly charged protactinium oxides \ce{PaO^{2+}} and \ce{PaO2^{2+}} \cite{Santos2006} leading to particular theoretical interest of the monoxide species \cite{kovacs:2011, zulch:2022, zulch:2023}. 
These studies collectively established the gas-phase chemistry of actinide oxides, but all rely on in-trap oxidation chemistry, and no charge states beyond $z = 2$ have been observed for any actinide monoxide or dioxide molecular ions.  In contrast, fluoride analogues have been produced in higher charge states, including \ce{UF^{3+}}  \cite{Schröder1999}, \ce{ThF^{3+}} \cite{Stricker2025} and the lanthanide molecule \ce{LaF^3+} \cite{Franzreb2004}.

Complementary evidence for multiply charged oxide molecules comes from laser-pulsed atom probe tomography (APT) and energetic oxygen sputtering experiments.
In laser-pulsed APT, doubly charged molecular oxide ions such as \ce{CeO_{1,2}^{2+}} and \ce{HfO_{1,2}^{2+}} \cite{Kirchhofer2013}, as well as actinide species including \ce{ThO_{1,2}^{2+}} \cite{Fougerouse2018, Fougerouse2020} and \ce{UO2^{2+}} \cite{Meisenkothen2020}, are emitted directly from bulk oxides by field evaporation.
More generally, field-ion microscopy and atom-probe studies have long established that extreme surface electric fields can generate multiply charged atomic and molecular ions for a wide range of metallic elements \cite{Mueller1974}.
However, access to charge states beyond 2+ for oxide molecules has so far been demonstrated only for selected d-block systems, including \ce{ReO^{3+}}, \ce{NbO^{3+}}, \ce{HfO^{3+}}, and \ce{NbO^{4+}}, produced via energetic oxygen-ion sputtering of metal targets \cite{Brites2011}.
While these experiments demonstrate that multiply charged oxide molecules can exist as metastable gas-phase species, the underlying field-evaporation and sputtering mechanisms do not provide controlled ion beams suitable for trapped-ion experiments or precision spectroscopy.

A new and more practical method to produce actinide molecules in higher charge states for trapped ion experiments is laser ablation. Here, doubly charged thorium monoxide cations (\ce{ThO^{2+}}) can be reliably produced via laser ablation of thorium metal foils \cite{Li2024, Claessens2025} and thorium fluoride molecules ThF$_x^{z+}$ (x = 0–3, z $\leq$ 3) from salt ablation \cite{Stricker2025}. Producing the molecular ions outside of the trap has several advantages, such as better vacuum conditions in the trap region, which is crucial for precision spectroscopy, and the possibility of mass selection of the ions of interest before trapping.
Uranium offers ideal experimental conditions through the long half-life of several isotopes, its availability and the fact that generating oxide molecules from metal foils is much more practical than the ablation of salt based targets, as metallic targets provide a homogeneous and well-defined material with a stable surface composition, whereas salt-based targets may exhibit compositional inhomogeneities and introduce additional chemical complexity (e.g., residual ligands or contaminants) during ablation. In this context, singly charged atomic and molecular uranium ions already play an important role in molecular, atomic, and nuclear spectroscopy and precision mass measurements studies \cite{Heaven2010, Battey2020, Han2022, Le2023, Polek2025, Kautz2021, Bubas2022, Bubas2022a, Raggio2024, Kromer2024}.

Our experiments were conducted using two complementary setups. At Johannes Gutenberg University Mainz (JGU), high-fluence laser ablation (HFLA) combined with a linear time-of-flight mass spectrometer enabled the production and identification of multiply charged uranium monoxide ions. At the University of Greifswald, high-precision multi-reflection time-of-flight (MR-TOF) mass spectrometry was employed to confirm ion identification and to exclude target contamination. 

The two setups are optimized for complementary experimental regimes. The production of multiply charged molecular ions requires sufficiently high laser fluence, which is achieved in the Mainz setup by tight focusing of the ablation laser. In contrast, the Greifswald MR-TOF setup operates at larger beam diameters and correspondingly lower effective fluences, limiting the formation of higher charge states ($z \geq 3+$). At higher fluences, the increased ion flux leads to detector limitations, requiring partial blocking of the ion beam to prevent saturation or damage of the channeltron detector, as demonstrated previously \cite{Fischer2025}. Consequently, the Greifswald setup is primarily used for high-resolution mass identification and purity studies, while the formation of highly charged species relies on the dedicated high-fluence conditions available in Mainz.

In this combined experimental and theoretical study, we establish multiply charged uranium oxide molecules as a previously inaccessible class of molecular ions for interdisciplinary investigations in physics and chemistry.
Using HFLA of a uranium metal foil, we directly produce and identify uranium monoxide ions \ce{UO^z+} in charge states $z=1$–4.
Among these, we observe the open shell ion \ce{UO^3+}, which relativistic density functional theory identifies as a promising candidate for precision tests of fundamental symmetries due to its comparatively simple electronic structure.
We further report the formation of the \ce{UO^4+} cation, which progresses into a largely unexplored regime of actinide chemistry at the edge of Coulomb instability.
Our calculations suggest that, unlike previously investigated quadruply charged molecular ions, \ce{UO^4+} exhibits metastability against Coulomb explosion and may therefore support bound vibrational states accessible to spectroscopic interrogation.

The employed approach provides a practical and readily reproducible route to the generation of isolated multiply charged actinide oxide ions directly from metal targets, suitable for trapping and/or spectroscopic experiments.

\section{Experimental and Computational Methods}

\subsection{Uranium metal foils}
\begin{figure}[!ht]
\includegraphics[width=0.9\linewidth]{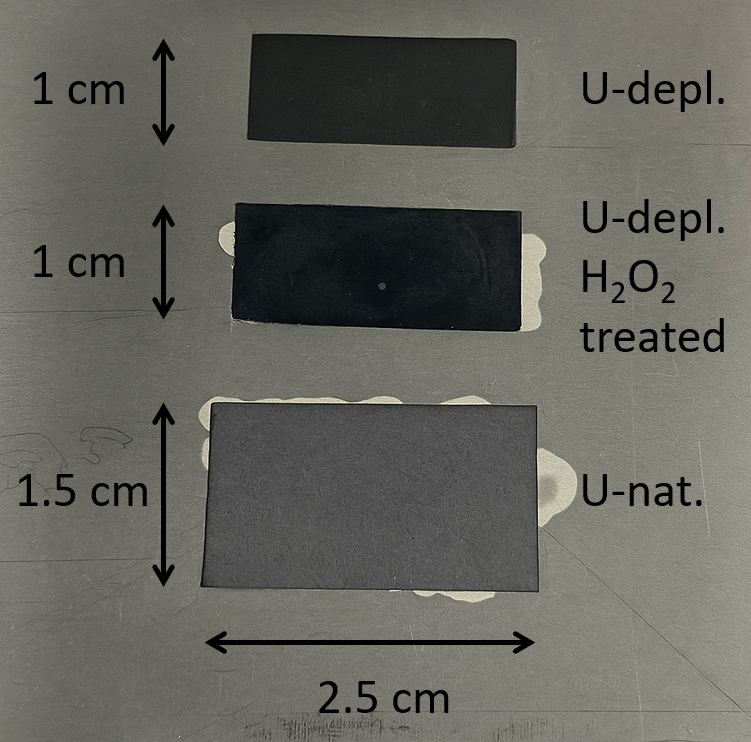}
\caption{Image of the ablated uranium foils, each 0.1 mm in thickness, on the Greifswald sample holder. Top: depleted uranium (478 mg; 7.17 kBq); middle: depleted uranium pretreated with hydrogen peroxide (478 mg; 7.17 kBq); bottom: natural uranium (717 mg; 17.90 kBq).}
\label{fig:targets}
\end{figure}

Three different uranium foils were used as target materials: depleted uranium, depleted uranium pretreated with hydrogen peroxide, and natural uranium (see Fig.~\ref{fig:targets}). 
Both the depleted and the pretreated depleted foils were cut from a single piece of metallic depleted uranium foil purchased in 2024 from Manufacturing Science Corporation (Oak Ridge, Tennessee, USA). 
The manufacturer-certified purity of the original foil is at least 99\% (see Supplemental Material \cite{SM} for details). 
The hydrogen-peroxide–treated sample was prepared by oxidizing a section of the original depleted foil in \ce{H2O2} for one hour to increase its surface oxidation state.
The natural uranium foil was originally supplied to the University of Marburg in 2014, transferred to Johannes Gutenberg University Mainz in 2024, and subsequently provided to the University of Greifswald for the laser ablation experiments. 
No manufacturer certificate was available for the natural uranium foil; its isotopic composition and purity were therefore analyzed experimentally in Sec.~\ref{LM}. 
All foils had a nominal thickness of 0.1~mm.

\subsection{\label{MR}Low-fluence ablation}

The MR-TOF device, described in \cite{Giesel2024, Fischer2024, Fischer2025}, was employed for ultra-precise mass analysis of the three different uranium foils (see figure \ref{fig:targets}). Ablation was driven by nanosecond pulses from a Litron TRLi DPSS 170-100 laser, operated at a wavelength of 532 nm and a repetition rate of 10 Hz. The laser spot diameter on the target was approximately 2 mm. For the present measurements, the pulse energy was varied between approximately 0.5 and 2 mJ, depending on the desired ion production rates. The targets (see Fig. \ref{fig:targets}) were glued to the sample holder and translated via a linear feed-through, so different regions of the foils could be irradiated without breaking the vacuum.
The holder was biased to $+2$ kV to accelerate cations towards the MR-TOF analyzer. After passing the analyzer, the ions were detected with a channeltron detector (DeTech 402AH).
The detector's geometry is allowing access for a photo-excitation laser. 
The ions’ kinetic energy exceeds all applied MR-TOF potentials. Thus, without a capture puls, the ions traverse the analyzer and reach the downstream detector in order of increasing mass-to-charge ratio. The resulting “single-path” spectrum, analogous to that of a linear TOF mass spectrometer, exhibits a mass resolving power $R = t / (2 \Delta t)$ on the order of 100.
For high-resolution, precision mass measurements, the ions are confined between the opposing electrostatic mirror potentials of the analyzer by applying an “in-trap” potential lift \cite{Wolf2012} to reduce their total energy. The storage greatly increases their flight time, while $\Delta t$ remains small due to the compensatory effect of the reflecting potentials: faster ions with higher kinetic energies penetrate more deeply into the mirror potential, thereby incurring longer flight paths. The high-resolution mass resolving power of the MR-TOF device depends on the number of ion revolutions and was typically set to $R \gtrsim 10^5$ in the present measurements.

\subsection{High-fluence laser ablation}

High-fluence laser ablation was performed with a linear time of flight mass spectrometer \cite{Stricker2025}. A piece of the depleted uranium foil (see figure \ref{fig:targets}), which was used beforehand in the MR-TOF setup, was exposed to high laser fluences in an ultra high vacuum ($< 10^{-8}$ mbar) setup for the production of molecular actinide ions in different charge states. The uranium target was 4 mm in diameter with a thickness of 0.1 mm, weighing 24 mg, and had an activity of 300 Bq. 

For laser ablation the target foil was glued to a Marcor-ceramic road with silver glue (Acheson 11415) and inserted into the TOF mass spectrometer. A Coherent FLARE NX71 515-0.6-2 laser, was operated at a wavelength of $\lambda = 515$ nm with a pulse duration of $1.3(2)$ ns and an energy output of  $E =300(15)$ µJ. The beam was focused to a diameter of $70(7)$ µm and a repetition rate of $1 - 10$ Hz was employed. The ablated cations were extracted by a constant voltage of 500 V applied to the extraction electrode, while the target itself was not electrically biased. The ions are accelerated by the resulting electric field immediately after ablation. The resulting spectrum of the linear TOF mass spectrometer exhibits a mass resolving power $R = t / (2 \Delta t)$ on the order of 150, where $t$ is the ion flight time and $\Delta t$ is the full width at half maximum (FWHM) of the arrival time distribution. It is limited primarily by the ions’ kinetic-energy spread.

\subsection{Data acquisition}

The data acquisition employed in the MR-TOF and HFLA measurements reflects their different experimental objectives.
In the MR-TOF spectrometer, ion counts are accumulated over many cycles in order to obtain absolute ion yields and enable a quantitative comparison between different atomic and molecular species, following established MR-TOF analysis procedures \cite{Fischer2018}.
In contrast, the HFLA spectra are averaged over repeated laser shots, with the primary focus on identifying and optimizing production conditions that maximize the temporally stable yield of specific molecular ions \cite{Stricker2025}.
This approach prioritizes reproducible ion generation under well-defined source parameters, which is essential for the subsequent extraction, mass selection, and capture of the ions in trapping experiments rather than for absolute yield determination.
While these two acquisition schemes are inherently different and not directly comparable on a quantitative level, they are ideally complementary for the present study.

In contrast to the MR-TOF measurements, where individual ions are detected and counted, the HFLA-based TOF spectra are recorded as a signal at the MCP detector, corresponding to the integrated charge of many ions. Together with intrinsic shot-to-shot fluctuations of the ablation process, this limits the direct quantitative comparison of ion yields, and the spectra are therefore primarily used for species identification and the analysis of qualitative trends.

\subsection{Computational details}
The DFT calculations were performed with a modified version
\cite{gaul:2020, zulch:2022, bruck:2023, colombojofre:2022} of a two-component program \cite{wullen:2010} based
on Turbomole \cite{ahlrichs:1989}. We employed the exchange correlation
functional by Perdew, Burke and Ernzerhof (PBE) \cite{perdew:1996} in its hybrid
versions with 25\,\% (PBE0) \cite{adamo1999} and 50\,\% Fock exchange (PBE50) \cite{bernard:2012}. Calculations of molecular properties were performed within the hybrid
local density approximation (LDA) using the X$\alpha$ exchange
functional \cite{dirac:1930,slater:1951} and the VWN-5 correlation
functional \cite{vosko:1980} with 50\,\% Fock exchange by Becke (BHandH)
\cite{becke:1993}. This functional has proven to be accurate for computation of $\mathcal{P,T}$-odd properties because results from more sophisticated methods such as
Coupled Cluster approaches lie usually between HF and LDA
\cite{gaul:2017,gaul:2020,gaul:2020a,gaul:2024,gaul:2024a}.

All calculations 
were performed within a quasi-relativistic complex generalized Kohn-Sham (cGKS)
framework including relativistic effects on the two-component zeroth order regular approximation
(2c-ZORA) level. 2c-ZORA was employed with a damped model potential to alleviate the gauge dependence
\cite{wullen:1998,liu:2002}. We employed atom-centered
Gaussian basis functions using the core-valence basis set of triple-$\zeta$
quality by Dyall (dyall.cv3z) \cite{dyall:2002, dyall:2006}. On uranium we used additionally double augmentation of the basis set (d-aug-dyall.cv3z). For calculations of $P,T$-odd properties we augmented the basis on U with an additional set of 12 s and 6 p functions for the description of the wave function within the nucleus. These additional sets of
steep functions were composed as an even-tempered series starting at
$10^{9}\,a_{0}^{-2}$ and progressing by division by $1.5$ for s
functions and $2.5$ for p functions.

The nuclear charge density distribution was
modeled as a normalized spherical Gaussian 
$\varrho_K \left( \vec{r} \right) = \frac{\zeta_K^{3/2}}{\pi ^{3/2}}
\text{e}^{-\zeta_K \left| \vec{r} - \vec{r}_K \right| ^2}$ with $\zeta_K =
\frac{3}{2 r^2 _{\text{nuc},K}}$. The root-mean-square radius
$r_{\text{nuc},K}$ was chosen as suggested by Visscher and Dyall employing the 
isotopes $^{16}$O and $^{238}$U \cite{visscher:1997}. Electronic densities were
converged until the change of the total energy between two consecutive cycles in
the self-consistent field procedure was below $10^{-10}\,E_\mathrm{h}$.
Molecular structures were optimized until the change of the norm of the 
gradient with respect to nuclear displacements was below
$10^{-3}\,E_\mathrm{h}/a_0$ and the change of the total energy was below
$10^{-6}\,E_\mathrm{h}$. The electronic ground state was found by fixing the occupation pattern to a specified electronic configuration. This was achieved by adjusting occupation numbers according to the maximum
overlap with the initial determinant [maximum overlap method (MOM)] \cite{gilbert:2008,barca:2018}. If
the change of the differential density matrix norm was below $10^{-3}$ between two consecutive
self-consistent field cycles, the maximum overlap with the determinant of the previous cycle was 
used. Although it is not guaranteed to find the global minimum, all angular momentum quantum numbers of lowest electronic states found in this work are consistent with known literature. The obtained electronic states were
characterized by computing the reduced total electronic angular momentum projection on the molecular axis (or in case of atoms on the $z$-axis) $\Omega=\vec{J}_\mathrm{e}\cdot\vec{\lambda}$ with $\vec{J}_\mathrm{e}=\vec{L}+\vec{S}$, where $\vec{L}$, $\vec{S}$ are the electronic orbital and spin angular momenta respectively and $\vec{\lambda}$ is the unit vector pointing from U to O. In addition we computed the expectation value of $\hat{S}^2$, $S(S+1)$. Although $L_z$, $S_z$ and $S$ are not good quantum numbers in a
relativistic framework, they can be used to estimate the composition of the 
spin symmetry-broken determinant from configuration state functions of non-relativistic
symmetry and give an approximate non-relativistic term symbol (see also Refs. \cite{zulch:2022,zulch:2023,Stricker2025}). For reproducibility we list all individual angular momenta and total energies in the Supplemental Material \cite{SM}. For the optimized electronic states we computed the distribution of electrons over the atoms using a Mulliken population analysis. 

Dissociation energies at room temperature were computed with thermodynamic corrections to molecular energies from harmonic vibrational frequencies with the module called ‘‘freeh’’ of the program package Turbomole for standard conditions (temperature 298.15 K and pressure 1 hPa)  assuming molecules are classical rigid rotors and harmonic oscillators. Atoms were treated as ideal gas assuming the entropy follows the Sackur--Tetrode equation \cite{Tedrode1912,Sackur1913}. For 0 K and 0 hPa we corrected all molecular energies by the zero-point vibrational (ZPV) energy. Dissociation energies without thermodynamic corrections and corresponding thermodynamic corrections can be found in the Supplemental Material \cite{SM}.

\section{Results and discussion}

\subsection{High precision MR-TOF analysis of uranium molecules}

\subsubsection{\label{LM}Characterization of uranium metal foils}

\begin{figure}[!ht]
\includegraphics[width=0.95\linewidth]{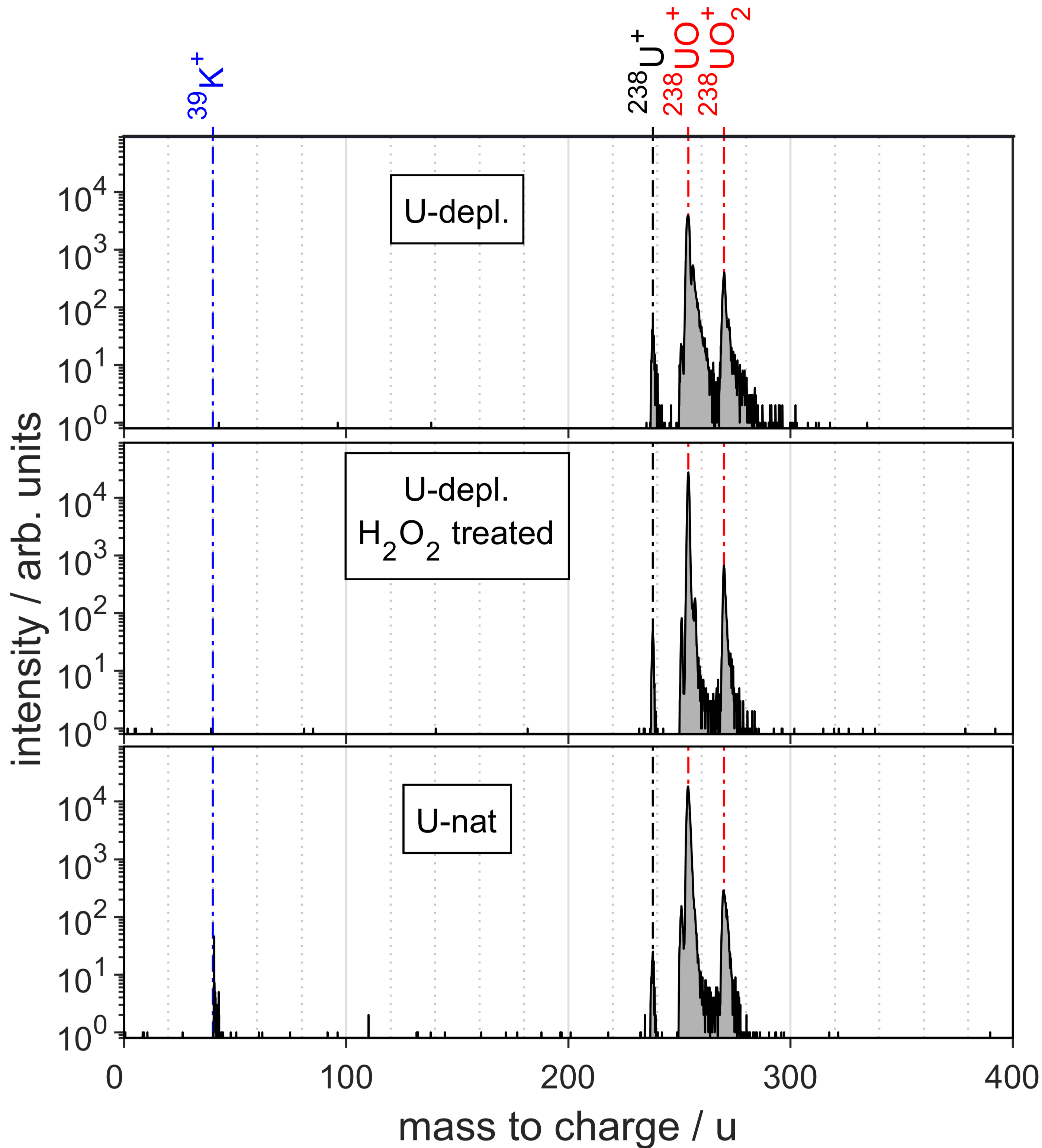}
\caption{Comparison of low-mass contaminants from laser ablation of depleted, \ce{H2O2}-treated, and natural uranium foils.}
\label{fig:cont}
\end{figure}

Figure \ref{fig:cont} shows single-path spectra produced at the Greifswald setup of the three uranium foils by laser ablation in the mass range up to 400~u for cations at low laser fluences. This range contains both low-mass contaminant and uranium-containing molecules with a single $^{235}$U or $^{238}$U atom.  
The laser pulse energy was adjusted in all measurements to avoid detector saturation on the most intense signals (at pulse energies of 1.0 mJ). To resolve low-abundance species, spectra were also accumulated with varying numbers of ablation shots. Absolute production rates cannot be compared directly, but the relative abundances of different ions remain consistent, indicating stable target conditions throughout.

The isotope ratios were determined via TOF measurements in Greifswald to $0.20(1)\%$ $^{235}$U for the depleted uranium foils and $0.75(7)\%$ $^{235}$U for the natural uranium foil, consistent with the declared specifications. For all uranium foils only \ce{K+} was observed as a contaminate species and only in trace amounts (Fig. \ref{fig:cont}). No additional species were detected in the mass range between potassium and uranium, confirming the absence of significant low-mass impurities and validating the charge-state analysis presented in Sec. \ref{charge}. Complementary, the absence of low-mass contaminants was confirmed by independent ICP-MS measurements (see \cite{SM}). In addition, no indication of additional low-mass species is observed at higher laser fluences, where the spectra are dominated by the formation of higher charge states and larger molecular aggregates.

\subsubsection{\label{cat}Uranium-containing compounds} 

\begin{figure*}[!ht]
\includegraphics[width=1\linewidth]{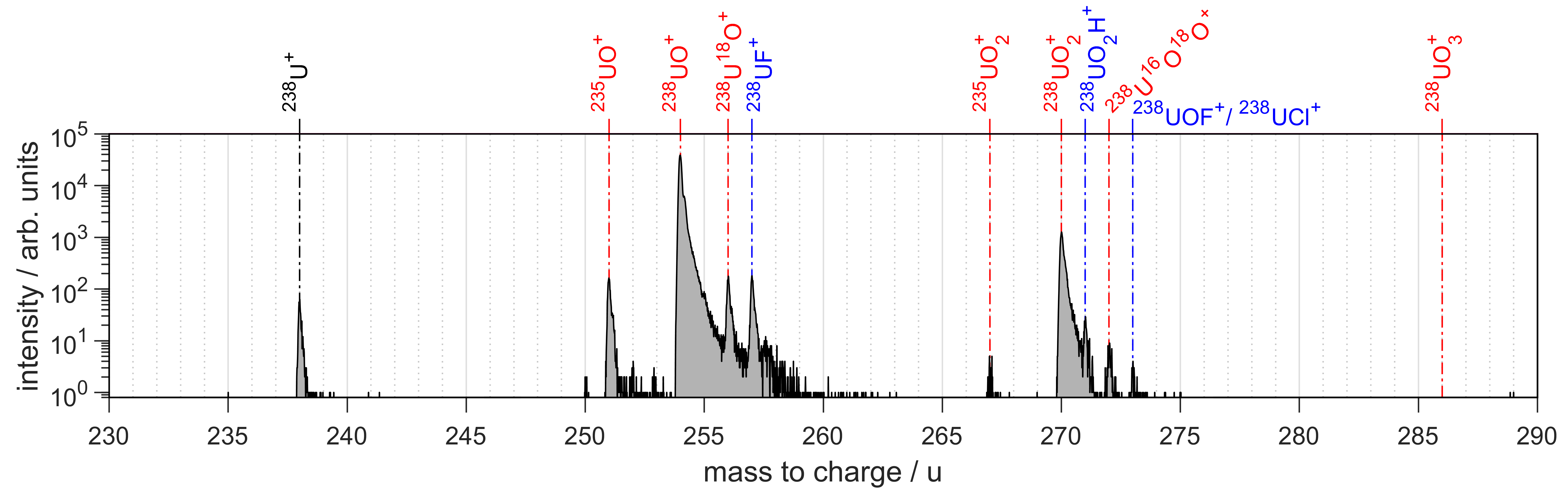}
\includegraphics[width=1\linewidth]{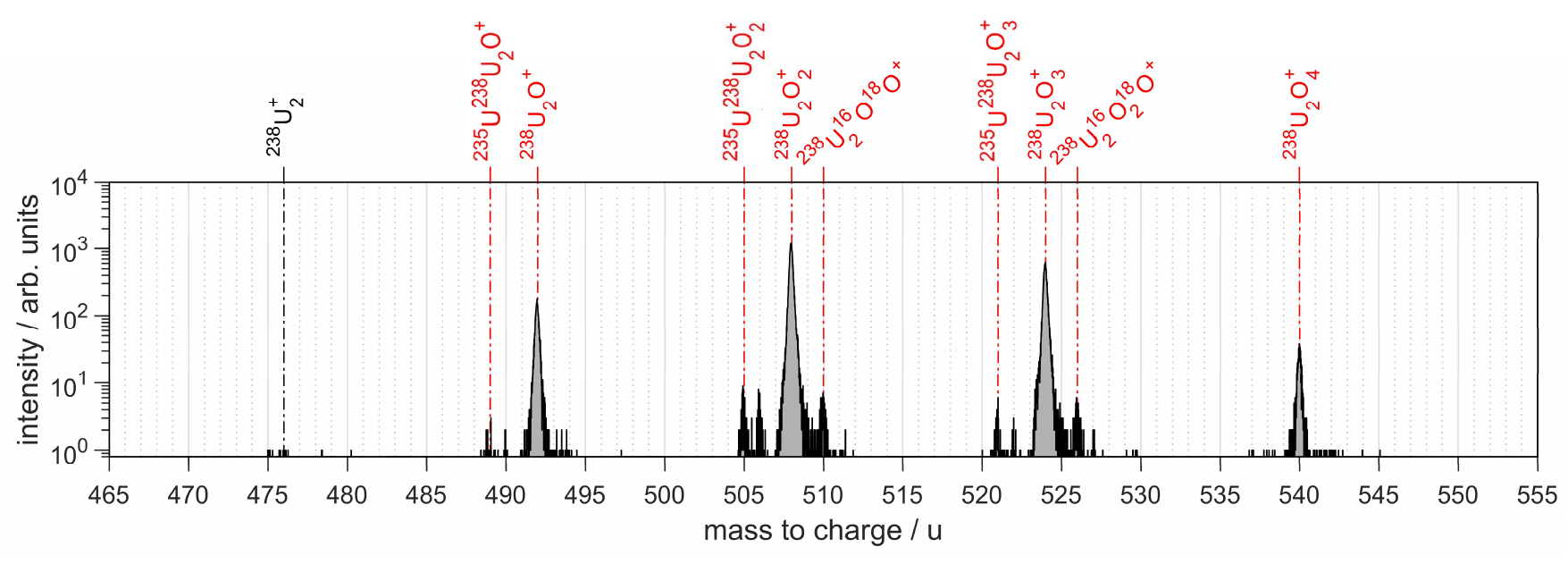}
\caption{Mass spectra of uranium-based species. The upper spectrum in the mass range of compounds containing one uranium atom is recorded at 10 revolutions in the MR-TOF analyzer with 1.0-mJ ablation pulse energy. The lower spectrum for compounds with two uranium atoms is recorded at 22 revolutions at the same pulse energy.}
\label{fig:cat}
\end{figure*}

The uranium species observed from the depleted uranium foil are shown in Fig.~\ref{fig:cat}. The spectra are dominated by oxygenated uranium species, while purely atomic uranium ions contribute only negligibly. In particular, \ce{UO+} emerges as the most abundant species, whereas multiply charged oxides such as \ce{UO3+} and \ce{UO4+} are not observed under the present conditions. Weak additional peaks around \(m/q \approx \) 252, 506, 523, and 527 u are tentatively attributed to low-abundance molecular uranium species. Several weak unidentified peaks exhibit a systematic mass spacing compatible with sequential oxygen addition following a possible nitrogen-containing precursor species. Since these signals were not individually confirmed in high-resolution MR-TOF mode, they are not assigned definitively

At elevated pulse energies, diuranium ions are observed, but only in the form of oxide-containing molecules (\ce{U2O_{1-4}+}), while the yield of bare diuranium ions remains extremely low. Notably, no evidence for uranium clusters beyond the diuranium stage is found, in stark contrast to laser ablation of metallic thorium foil, which readily produces metallic clusters \cite{Fischer2025}. This highlights a fundamental difference in the cluster-forming propensity of uranium versus thorium under nanosecond ablation conditions. These results suggest that uranium clustering is strongly suppressed, and that the observed dimers are stabilized by oxygen bridges rather than by direct U–U bonding. The complete absence of triuranium oxides (\ce{U_xO_y}  with $x\geq3$) and of pure metallic clusters (\ce{U_x}  with $x\geq3$) across all tested foil types (depleted, pretreated/oxidized, and natural) demonstrates that the ion distribution is remarkably robust against variations in the chemical state of the target, emphasizing that laser–plasma parameters play the decisive role in determining the accessible species.

\subsection{\label{charge}Production of multiply charged atomic and uranium monoxide ions}

\begin{figure}[!ht]
\includegraphics[width=1\linewidth, height=1\textheight, keepaspectratio]{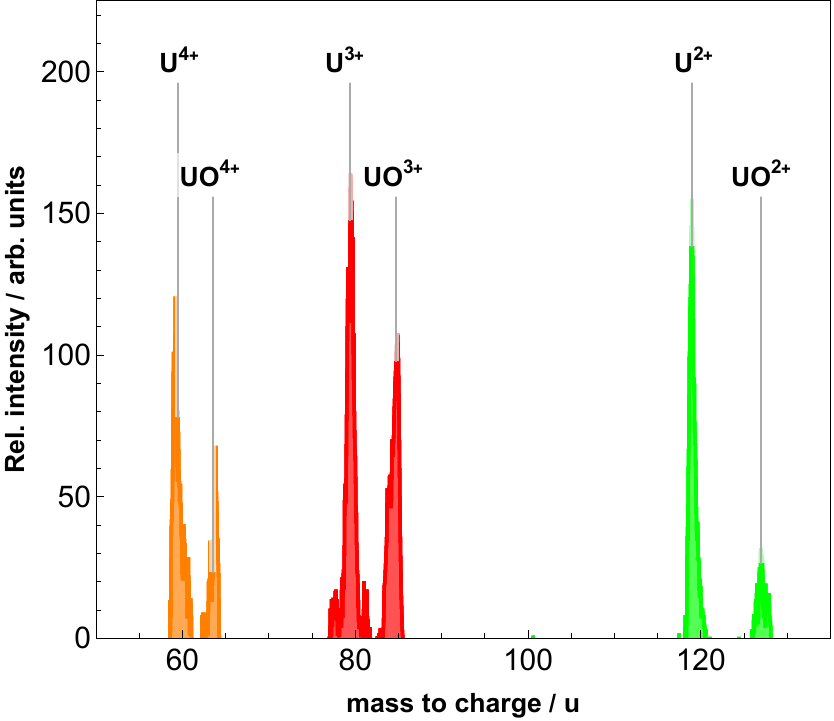}
\caption{Spectra of atomic and monoxide uranium ions in charge states 2+ (green), 3+ (red) and 4+ (orange). Laser fluence was varied in the range from 3.40 to 4.70 J$\cdot$cm$^{-2}$ and five recorded spectra were summed. The laser fluences for given species (\ce{U^{z+}} and \ce{UO_{0-1}^{z+}} with $z = 2,3, 4$) are listed in Table \ref{tab:Charge}.} 
\label{fig:Uohne}
\end{figure}

The generation of uranium molecular ions in charge states $n>1$ was investigated focusing on the dependence of charge-state formation on laser spot size and fluence. Importantly, no contaminants were detected in the mass-over-charge ($m/q$) range between 40 and $235\,u/q$ (Sec.~\ref{LM}), such that all observed signals in this range can be unambiguously assigned to uranium-containing ions in higher charge states. 

\begin{table}
\caption{Production of atomic and oxide-containing U ions in charge states 2+ to 4+. The table shows the fluence ( "$\Psi$" in J$\cdot$cm$^{-2}$) ranges in which the respective species dominates.}
\label{tab:Charge}
\begin{ruledtabular}
\begin{tabular}{c|cccc}
Ion species & 4+ & 3+ & 2+& \\
\hline 
\ce{U} &$4.70\pm0.12$ &  $4.30\pm0.11$ & $3.70\pm0.09$ &\\ 
\ce{UO} &$4.70\pm0.12$ &  $4.10\pm0.10$ & $3.40\pm0.08$ &\\ 
\end{tabular}
\end{ruledtabular}
\end{table}

The production of both atomic uranium (\ce{U^z+}) and uranium monoxide (\ce{UO^z+}) ions requires laser fluences $\Psi$ above $\sim$3.00\,J$\cdot$cm$^{-2}$. A clear transition occurs at $\Psi \approx 4.70\pm0.12$\,J$\cdot$cm$^{-2}$, where ionization competes with bond dissociation. At this threshold, the simultaneous appearance of \ce{U^4+} and \ce{UO^4+} marks the threshold for formation of the 4+ charge state. The peak shapes in Fig. \ref{charge} shows moderate variations between different species and charge states. These differences can be attributed primarily to species-dependent variations in the initial phase-space distribution following the high fluence laser ablation. In particular, differences in kinetic energy spread and formation dynamics within the ablation plume can lead to slight variations in the arrival-time distributions.
The ion signal was detected with an MCP detector. From the integrated signal of the UO$^{3+}$ peak, with a 50 $\Omega$ termination and a MCP gain of $10^6$, we estimate an ion yield on the order of approximately $10^2$ to $10^4$ ions per laser shot per peak. This represents a lower bound due to finite transport and detection efficiencies. Such ion yields are compatible with trapped-ion experiments, which typically rely on the preparation of a small number of ions and repeated loading cycles.

The process is reproducible, underscoring that the laser and extraction conditions are decisive rather than the detailed chemical state of the foil. This robustness opens a straightforward pathway for physics and chemistry laboratories worldwide to reliably generate multiply charged actinide molecules for spectroscopic applications and precision studies.

No evidence was found for atomic or oxidic species in charge states $n\geq5$, fully consistent with our quantum-chemical analysis of thermodynamic stability. The calculations suggest that \ce{UO^4+} is metastable, with a barrier of significantly less than 1\,eV towards Coulomb explosion (see \ref{QC}). This indicates that the quadruply charged monoxide already represents a limiting case: any further electron removal would probably lead to immediate bond cleavage, rather than formation of a stable \ce{UO^5+} species. 
While the generation of atomic \ce{U^5+} at higher fluences remains plausible, the existence of molecular uranium oxides beyond \ce{UO^4+} would require stabilization mechanisms not captured by the present experiments or theoretical framework. Identifying or excluding such mechanisms represents a critical challenge for future work, as it will determine whether a fundamental energetic boundary defines the upper limit of actinide molecular ion chemistry.

\subsection{\label{QC}Quantum chemical study of thermodynamic stability}
\subsubsection{Accuracy of the employed methodology}
We benchmarked the performance of employed exchange-correlation functionals by computing the first to sixth ionization energies of U as well as the ionization energy of oxygen (\ref{tab:partialcharges}). 
We find an excellent agreement for the of 2c-ZORA-cGKS-PBE0 results, reflected in relative deviations below 5\,\% to literature values of ionization energies of uranium. An exception is the fourth ionization energy where we find a deviation of 10\,\%.
The ionization energy of oxygen is in perfect agreement with literature (deviation below 1\%). The accuracy of results obtained with the PBE50 functional is generally lower. Deviations are, however, still below 7\,\% except for the fourth ionization energy of uranium (11\,\%).

\begin{table}
\caption{Approximate non-relativistic electronic ground state configurations of different charge states of uranium and oxygen and corresponding ionization energies as computed at the levels of 2c-ZORA-cGKS-PBE50/d-aug-dyall.cv3z and 2c-ZORA-cGKS-PBE0/d-aug-dyall.cv3z compared to experimental (exp) and theoretical (theo) data in the literature.}
\label{tab:partialcharges}
\begin{ruledtabular}
\begin{tabular}{llS[round-precision=2,round-mode=places]S[round-precision=2,round-mode=places]Sl}
Atom & Term& \multicolumn{2}{c}{$I/\mathrm{eV}$} & {$I_\mathrm{ref}/\mathrm{eV}$} & Ref.\\
&&{PBE50}&{PBE0}\\
\hline
\ce{O}     & $^{3}\mathrm{P}_2$   & 13.5036&13.6438& 13.618055(7)&exp\cite{Eriksson1968}\\
\ce{O+}    & $^4\mathrm{S}_{3/2}$ & \\
\ce{U}     & $^{5}\mathrm{L}_{6}$    & 6.28816 & 6.14166    &   6.19405(6)& exp\cite{Coste1982}    \\
\ce{U+}    & $^{4}\mathrm{I}_{9/2}$    & 12.0659 &  11.5454 &   11.6(4)   & exp\cite{Blaise1992}  \\
\ce{U^2+}  & $^{5}\mathrm{I}_{4}$      & 18.5193 &  19.4178 &   19.8(3)   & exp\cite{Blaise1992}  \\
\ce{U^3+}  & $^{4}\mathrm{I}_{9/2}$ & 32.7802  &  33.0256   &   36.7(10)  & exp\cite{Blaise1992}  \\
\ce{U^4+}  & $^{3}\mathrm{H}_{4}$ & 47.5639  &  47.7532     &   46.0(19)  & theo\cite{Rodrigues2004}  \\
\ce{U^5+}  & $^{2}\mathrm{F}_{5/2}$ & 63.2078  &  63.4576   &   62.0(16)  & theo\cite{Blaise1992}  \\
\ce{U^6+}  & $^{1}\mathrm{S}_{0}$ &   &    & \\
\end{tabular}
\end{ruledtabular}
\end{table}

Considering molecular species, approximate term symbols for electronic ground states of \ce{UO^z+} with $z=0,1,2$ are in agreement with previous theoretical studies \cite{kovacs:2015}. The determined electronic ground state of \ce{UO^3+} is equal to the electronic ground states in the isoelectronic species \ce{PaO^2+} \cite{kovacs:2013,zulch:2022,zulch:2023} and \ce{PaF^3+} \cite{zulch:2022}. Our results for equilibrium bond length, harmonic vibrational wavenumbers and dissociation energies for \ce{UO^z+} with $z=0,1,2$ are compared in Table \ref{tab:partialcharges} with previous calculations and experimental values. All results are in good agreement with previous literature. Bond length are slightly underestimated by 1\,\% and harmonic vibrational wavenumbers are overestimated by 3\,\%, respectively, for the PBE0 functional. Deviations are slightly larger for PBE50. This indicates, that the present calculations may slightly overestimate the overall bond strength in \ce{UO^z+}. This is also reflected in an overestimation of the dissociation energies of \ce{UO} and \ce{UO^2+} by roughly 3\,\%. Although the PBE50 functional predicts slightly shorter bonds and larger harmonic vibrational wavenumbers, predicted dissociation energies are generally lower than for PBE0. Overall we conclude that the present methodology is suitable for determining the thermodynamic stability of uranium monoxide molecules with an accuracy of several percent. 

\subsubsection{Thermodynamic stability of \ce{UO^3+}}

In Table \ref{tab:partialcharges} we summarize all results for \ce{UO^z+} with $z=0,1,2,3,4$. Our calculations show that \ce{UO^3+} is meta-stable against Coulomb explosions. Although, the Coulomb explosion channel lies 2.6\,eV below the energetic minimum of bound \ce{UO^3+}, we could not find an avoided crossing or spin-orbit mixture of the dissociative potential up to interatomic distances of $2.5\,\AA$. This is in agreement with a rough diabatic approximation of the Coulomb explosion channel: for an asymptotically flat behavior of the diabatic repulsive potential we would expect that at the crossing point of the potentials $R_\mathrm{crit}$ is at $D_{0,\ce{O+U^3+}}=D_{0,\ce{O+U^3+}}+\frac{Q_1Q_2}{4\pi\epsilon_0R_\mathrm{crit}}$. It follows $R_\mathrm{crit}\sim\frac{28.8\,\mathrm{eV}}{5.0\,\mathrm{to}\,5.8\,\mathrm{eV}}\,\text{\AA}\sim5.0\,\mathrm{to}\,5.7\,\text{\AA}$. We therefore expect a barrier of considerable width ($\Delta R\gg 0.85\,\AA$) and height $\Delta E=1.90\,\mathrm{to}\,3.16\,\mathrm{eV}$. In a WKB approximation and assuming a simple step function for the barrier shape this would lead to extremely low tunneling dissociation rates $10^{13}\mathrm{exp}\left(-2\times1.6\,a_0\times\sqrt{2\times27343\,m_\mathrm{e}\times0.07\,\frac{\hbar^2}{m_\mathrm{e}a_0^2}}/\hbar\right)\,\mathrm{Hz}\sim10^{-73}\,\mathrm{Hz}$, where $10^{13}\,\mathrm{Hz}$ is the fundamental vibrational frequency in \ce{UO^3+} and $\sim15\,\mathrm{amu}\approx27343\,m_\mathrm{e}$ is the reduced mass. Therefore, we expect the electronic ground state to be observationally stable. The stability against Coulomb explosion is reflected in the computed Mulliken partial charges. In \ce{UO^3+} the positive charge is located almost entirely on uranium 

\begin{figure}[!ht]
\includegraphics[width=\linewidth]{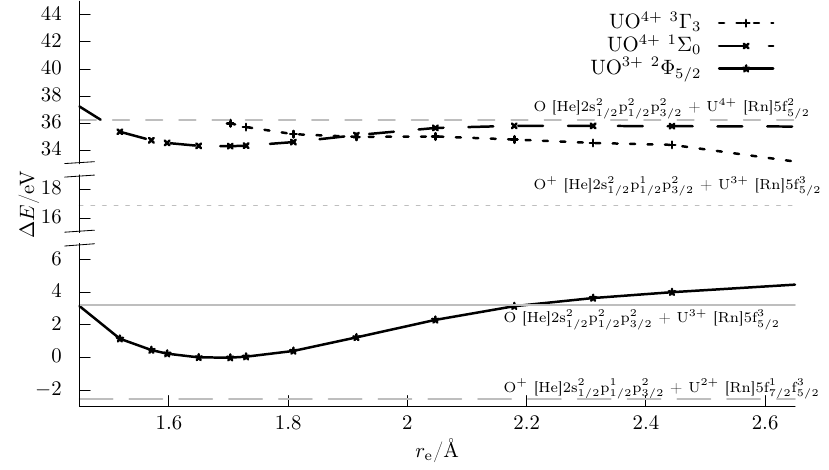}
\caption{Relevant excerpts of the potential energy curves of \ce{UO^3+} and \ce{UO^4+} for estimating  dissociation energies at the level of 2c-ZORA-cGKS-PBE0/d-aug-dyall.cv3z. Dissociation energies computed from separate atoms are shown as gray horizontal lines. The crossing of the triplet and singlet curves of \ce{UO^4+} happens at about 1.9\,\AA. Lines are shown to guide the eye. Because of an admixture of quartet states the potential curve of the ${}^2\Phi_{5/2}$ in \ce{UO^3+} lies between 2.2\,\AA\  and 2.6\,\AA\  above the dissociation limit.\footnote{Although not explicitly computed in the present work, we expect the quartet state ${}^4\Gamma_{5/2}$ corresponding to the Coulomb explosion channel to lie below the ${}^2\Phi_{5/2}$ state for bond length larger than 2.2\,\AA. Moreover, for longer bond lengths than 2.6\,\AA, it is likely that the ${}^6H_{5/2}$ and ${}^8I_{5/2}$ states of \ce{UO^3+} subsequently are lower than the ${}^4\Gamma_{5/2}$ state. For \ce{UO^4+} the ${}^5H_{3}$ and ${}^7I_{3}$ states will probably subsequently lie lower than the ${}^4\Gamma_{3}$ one towards bond dissociation.}}
\label{fig:diss}
\end{figure}

\begin{figure*}[!ht]
\includegraphics[width=1\linewidth]{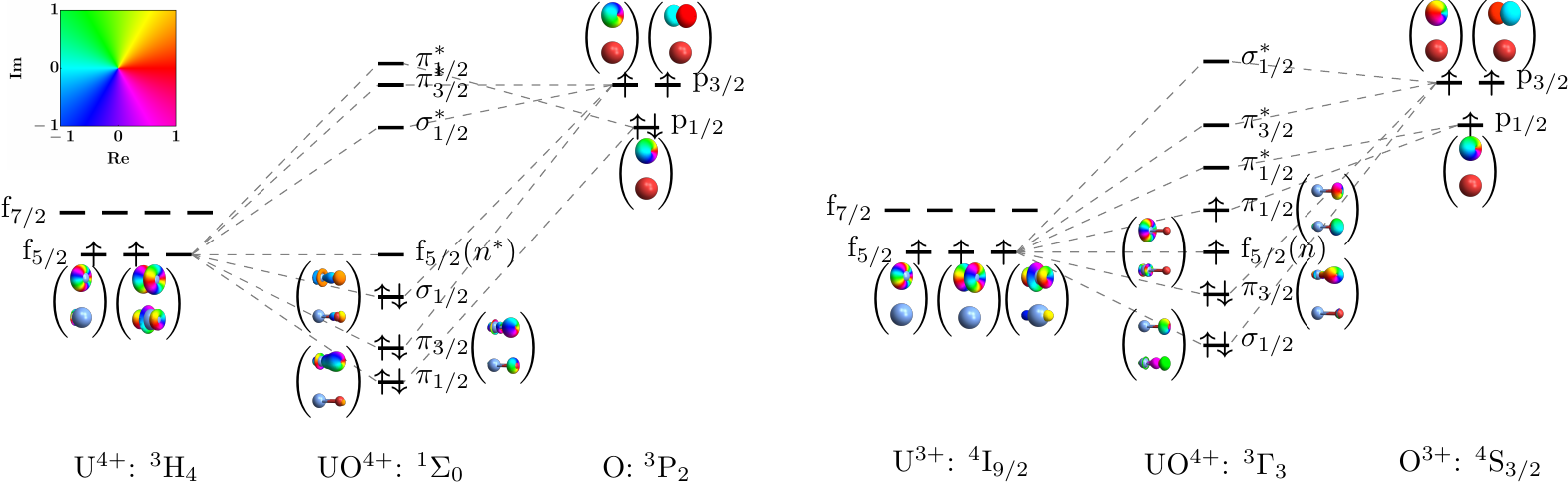}
\caption{Qualitative molecular orbital scheme for \ce{UO^4+} in the determined electronic ground state with a singlet configuration (left) and the triplet state at a bond length of $r_\mathrm{e}+\SI{0.5}{\bohr}$, which is connected to a Coulomb explosion into \ce{U^3+} and \ce{O^+}. Occupied Kohn-Sham spinors are visualized for a single approximate Kramers partner by calculating the amplitudes of spin-up and spin-down components on a three-dimensional grid. Subsequently the phase in the complex plane was mapped via the color code shown in the upper left square on the contour surface of the absolute value of the spinor using the program Mathematica \cite{Mathematica} version 14.3 incorporating a contour value of \SI{0.1}{\bohr^{-3/2}}. The U atom is represented with a light blue ball, the O atom with a red ball. The energetic order of spinors follows qualitatively the averaged Kohn-Sham spinor energies, their positioning is not quantitative. Note that the phase and orientation of the total angular momentum of each molecular orbitals is arbitrary. In a spin-orbit coupled framework the wave function is not an eigenstate of $\hat{\vec{L}}$ or $\hat{\vec{S}}$. Therefore, all term symbols are to be understood as approximate. Symmetry labels are chosen with respect to the largest contribution of eigenstates of $S$ to spin-orbit coupled states.}
\label{fig:mo}
\end{figure*}

\subsubsection{Thermodynamic stability of \ce{UO^4+}}
In contrast to \ce{UO^3+}, in \ce{UO^4+} a significant fraction of charge is found on oxygen ($+0.4\,e$).
The computed dissociation energies suggest that in \ce{UO^4+} other bonding mechanisms like covalent interactions or correlation effects overcome the electrostatic repulsion. This can be clarified with a dedicated analysis of the chemical bond in \ce{UO^4+}, which is, however, beyond the scope of the present study.
 
To estimate the barrier towards Coulomb explosion in \ce{UO^4+}, we computed the potential energy curve of the electronic ground state  (${}^1\Sigma_0$) dissociating in \ce{O} ($\mathrm{{}^3P_{2}:~[He]2s_{1/2}^2p_{1/2}^2p_{3/2}^2}$) and \ce{U^{4+}} ($\mathrm{{}^3H_{4}:~[Rn]5f_{5/2}^2}$) and the crossing point with the triplet state (${}^3\Gamma_3$) which dissociates to \ce{O+} ($\mathrm{{}^4S_{3/2}:~[He]2s_{1/2}^2p_{1/2}^1p_{3/2}^2}$) and \ce{U^3+} ($\mathrm{{}^4I_{9/2}:~[Rn]5f_{5/2}^3}$) at the level of PBE0 as shown in Figure \ref{fig:diss} (the corresponding figure for PBE50 is provided in the Supplemental Material \cite{SM}).
To illustrate the involved electronic configurations a qualitative molecular orbital scheme for the singlet state at equilibrium bond length and the triplet state close to the crossing point of the potential curves  is provided in Figure \ref{fig:mo}. 
We find a barrier of \SI{0.63}{\electronvolt} for a transition from the singlet to the triplet electronic state in \ce{UO^4+} with a crossing point being at $R_\mathrm{crit}\approx1.9\,\text{AA}$ (computed at the level of PBE0 including a ZPV correction of \SI{-0.06}{\electronvolt}, at the level of PBE50 we find \SI{0.40}{\electronvolt} with a crossing point at $R_\mathrm{crit}\approx1.8\,\text{AA}$). For estimating the lifetime of vibrational states we may assume again the rough WKB approximation with a step function: at the level of PBE0 we find $\Delta R=0.26\,\text{\AA}$ and $\Delta E=0.63\,\mathrm{eV}$ resulting in a dissociation rate through vibrational tunneling of $\nu_\mathrm{e}\mathrm{exp}\left(-2\times0.49\,a_0\times\sqrt{2\times27343\,m_\mathrm{e}\times0.023\,\frac{\hbar^2}{m_\mathrm{e}a_0^2}}/\hbar\right)\sim8\times10^{-16}\,\nu_\mathrm{e}$, suggesting a life-time of the vibrational ground state of 10s to 100 seconds. Therefore, within a harmonic approximation vibrational states with quantum numbers $v<6$ are quasi-bound. The exact form of the barrier and anharmonic effects may slightly alter these estimates. The large reduced mass in \ce{UO^z+} results in comparatively slow tunneling rates which may lead to long life-times of higher-lying vibrational states. In contrast to lighter metastable molecular anions \cite{schroder:2005}, these may therefore considerably exceed nanoseconds. If \ce{UO^4+} is formed predominantly by ionization of \ce{UO^3+}, the vibrational population of \ce{UO^4+} is expected to be governed by Franck--Condon factors between the vibrational states of \ce{UO^3+} and \ce{UO^4+}.
In this case, the resulting vibrational-state distribution may favor low-lying vibrational levels of \ce{UO^4+}.
Such a population bias is consistent with the comparatively high experimental signal intensity observed for \ce{UO^4+}. 
At the computational level of PBE50 we find a slightly less favorable situation ($\Delta R=0.11\,\text{\AA}$ and $\Delta E=0.40\,\mathrm{eV}$) resulting in a much shorter life-time of the vibrational ground state on the order of 10\,ns, which would contradict the experimental observation of \ce{UO^4+}. From comparison of properties of \ce{UO^z+}, for $z=0,1,2$ with previous studies we see that the large amount of Fock exchange in PBE50 systematically underestimates chemical bond strengths. We therefore consider the PBE0 results to be more reliable.
This analysis demonstrates that a detailed theoretical study of the excited state manifold with more sophisticated relativistic multi-reference methods will be crucial towards spectroscopic studies of metastable multiply charged actinide molecules.

\begin{table*}
\caption{Approximate non-relativistic electronic ground state configuration, equilibrium bond length $r_\mathrm{e}$, harmonic vibrational wavenumber $\tilde{\omega}_\mathrm{e}$, dissociation energies for the homolytic bond cleavage ($D_{T,\ce{O}+\ce{U^z+}}$) and the Coulomb explosion ($D_{T,\ce{O+}+\ce{U^{(n-1)}+}}$) channel at $T=298$ K, $p=1$ hPa and at $T=0$ K, $p=0$ hPa and Mulliken partial charges of \ce{UO^z+} with $z=0,1,2,3,4$ molecular ions computed at the levels of 2c-ZORA-cGKS-PBE50/d-aug-dyall.cv3z and 2c-ZORA-cGKS-PBE0/d-aug-dyall.cv3z for the optimized electronic states (see Sec. II.B and Supplemental Material \cite{SM} for details).}
\label{tab:partialcharges}
\begin{ruledtabular}
\begin{tabular}{lll
S[round-mode=places, round-precision=2]
S[round-mode=places, round-precision=0]
S[round-mode=places, round-precision=2]S[round-mode=places, round-precision=2]S[round-mode=places, round-precision=2]S[round-mode=places, round-precision=2]
S}
Molecule & Term& Method & \multicolumn{1}{c}{$r_\mathrm{e}/$\AA} & \multicolumn{1}{c}{$\tilde{\omega}_\mathrm{e}/\mathrm{cm}^{-1}$} & \multicolumn{1}{c}{$D_{0,\ce{O}+\ce{U^z+}}$} & \multicolumn{1}{c}{$D_{298,\ce{O}+\ce{U^z+}}$} & \multicolumn{1}{c}{$D_{0,\ce{O+}+\ce{U^{(n-1)}+}}$} & \multicolumn{1}{c}{$D_{298,\ce{O+}+\ce{U^{(n-1)}+}}$} &\multicolumn{1}{c}{$\delta_\mathrm{U}/e$}  \\
\cline{6-9}
&&&&&\multicolumn{4}{c}{eV}\\
\hline
\multirow{3}{*}{\ce{UO}}     & \multirow{3}{*}{$^{5}\mathrm{I}_{4}$}      
 &PBE50&  1.8142 &895.83  & 7.37669& 7.26954& {-} & {-} &0.58\\
&&PBE0 &  1.8244 &873.15  & 8.2451 & 8.13832& {-} & {-} &0.52\\
&&exp\cite{kaledin:1994,konings:2006}&1.8383(6)&911.9(2)& {-}& 7.84(13) \\
\\
\multirow{3}{*}{\ce{UO+}}    & \multirow{3}{*}{$^{4}\mathrm{I}_{9/2}$}    
& PBE50 &  1.7753 &971.70  & 7.78424& 7.67585& 14.9996& 14.8912&1.45 \\
&&PBE0 &   1.7873 &943.49  & 8.34156& 8.23361& 15.8437& 15.7358&1.39\\
&&exp\cite{goncharov:2006,gibson:2006}&1.801(5)&846.5(5)&  & 8.3(3) \\
\\
\multirow{3}{*}{\ce{UO^2+}}  & \multirow{3}{*}{$^{3}\mathrm{H}_{4}$} 
& PBE50&  1.7008 &1129.97 & 6.9509 & 6.84022& 8.38857& 8.27788&2.20 \\
&&PBE0&  1.7176 &1077.66 & 7.02792& 6.91774& 9.12633& 9.01615 &2.15\\
&&theo.\cite{kovacs:2011}/exp.\cite{gibson:2005}\footnote{Theoretical values for $r_\mathrm{e}$ and $\tilde{\omega}_\mathrm{e}$, experimental value for $D_{298,\ce{O}+\ce{U^z+}}$.}&1.728&1047.2 & & 7.1(6) \\
\\
\multirow{2}{*}{\ce{UO^3+}}  & \multirow{2}{*}{$^{2}\mathrm{\Phi}_{5/2}$} 
& PBE50&  1.6534 &1202.32 & 1.89718 & 1.78503&-3.1186 & -3.23075&2.93 \\
&&PBE0&  1.6828 &1096.00 & 3.15698& 3.04576&-2.61695& -2.72818 &2.87\\
\\
\multirow{1}{*}{\ce{UO^4+}}  & \multirow{1}{*}{$^{1}\mathrm{\Sigma}_{0}$} 
& PBE50& 1.6389 &1153.50 & 0.00974308 & -0.10282 & -19.2669& -19.3795&  3.63\\
&& PBE0& 1.6878 &984.11 &  1.9021& 1.79115 &-17.4796 &-17.5906 &  3.56\\
\end{tabular}
\end{ruledtabular}
\end{table*}

\subsection{Fundamental physics sensitivity of triply charged uranium monoxide}
\ce{UO^3+} is interesting for precision spectroscopy studies of fundamental physics as it features a comparatively simple electronic structure with a single valence electron, which is rarely found in actinide compounds. At the same time the uranium nucleus is a promising probe of fundamental interactions as it features several long-lived isotopes with deformed nuclei.

\subsubsection{Studying nuclear moments of uranium}
The \ce{UO^3+} molecule provides a promising platform for studying the nuclear moments of uranium isotopes. For example, we anticipate studies of the nuclear magnetization distribution, which has been observed in a molecule until now only once \cite{wilkins:2025}, and higher order nuclear moments like the nuclear magnetic octupole moment which remain unobserved in molecular spectroscopy. These are expected to be sufficiently large for observation in \ce{UO^3+} because (i) the long-lived ${}^{233}$U, ${}^{235}$U nuclei feature high angular momentum and octupole collectivity, (ii) the internal molecular fields are expected to exhibit a considerable relativistic enhancement in \ce{UO^3+}, and (iii) low-lying electronic states of \ce{UO^3+} have sufficiently large angular momenta for observing these higher order nuclear moments. While the present theoretical work focused on the most abundant odd-$A$ isotope $^{235}$U, the experimental approach is directly applicable to all sufficiently long-lived uranium isotopes, thereby enabling systematic studies of isotope-dependent nuclear properties in a molecular environment. 

\subsubsection{Possible routes to precision spectroscopy of triply charged uranium monoxide}
 In order to achieve precision spectroscopy of \ce{UO^3+} a necessary next step will be the spectroscopic characterization of the molecule.

For studying $CP$-violation, a measurement protocol as proposed in Ref.\,\cite{zhang:2023} is a potential route for in-trap precision spectroscopy of \ce{UO^3+}. In this type of measurement, high angular momentum in the science state is beneficial. The electronic ground state of \ce{UO^3+} would thous be suitable for an EDM measurement. However, a detailed combined theoretical and experimental study of the rovibronic structure of the molecule is required to make a final assessment.  In particular $\Omega$-doubling constants will be crucial for deciding on the exact measurement protocol.

Considering other types of precision spectroscopy, without detailed knowledge of the excited electronic states we cannot yet provide a specific protocol for measurement but we believe that quantum logic spectroscopy in an ion trap with for example \ce{Ca+} ions \cite{Stopp2019, GrootBerning2018, Li2021} provides an interesting avenue we want to pursue in future works.

\subsubsection{Sensitivity to nuclear $CP$-violation}
Uranium isotopes  feature well-deformed nuclei \cite{butler:2016}. The odd-$A$ isotopes ${}^{233}$U, ${}^{235}$U are expected to show octupole collectivity \cite{thompson:1976}, where $A$ is the nuclear mass number. Therefore, they may strongly enhance simultaneous violations of parity ($P$) and time-reversal symmetry ($T$), which manifests in $P,T$-odd nuclear Schiff moments \cite{flambaum:2025} or nuclear magnetic quadrupole moments (NMQMs) \cite{flambaum:2022}.  Such $P,T$-odd moments are an indirect signature to $CP$-violation \cite{khriplovich:1997}---a necessary ingredient for baryogenesis \cite{sakharov:1967}.

In Ref.\,\cite{flambaum:2020} the Schiff moment of ${}^{233}$U and ${}^{235}$U was estimated to be strongly enhanced due to octupole collectivity being roughly 2 or 3 times as large as in ${}^{225}$Ra, respectively.

To the best of our knowledge only one U-containing molecule was studies with respect to $P,T$-odd effects before: In a study of \ce{UF^3+}, a rich electronic structure was found which is probably impractical for the purpose of precision experiments \cite{zulch:2023}. In the following we evaluate the electronic structure sensitivity of \ce{UO^3+} to $P,T$-violation. According to our calculations \ce{UO^3+} has an electronic ground state similar to \ce{PaF^3+},  which has been predicted to exhibit favorable sensitivity for precision studies of fundamental physics \cite{zulch:2022,gaul:2024a,zulch:2025}, although its experimental realization and control remain to be demonstrated.

$P,T$-violation manifests itself in a permanent electric dipole moment (EDM) of a molecule. To compute the sensitivity of a molecular EDM to fundamental sources of $P,T$-violation, we consider possible contributions discussed in Refs.~\cite{gaul:2024,gaul:2024a,degenkolb:2024,athanasakis-kaklamanakis:2025}: We consider the tensor-pseudotensor $k_{\rm{T}}$, scalar-pseudoscalar $k_{\rm{s}}$, and pseudoscalar-scalar $k_{\rm{p}}$ nucleon-electron current interaction coupling constants, the electron EDM $d_{\rm{e}}$, the short-range neutron EDM $d^{\rm{sr}}_{\rm{n}}$, the isoscalar $\bar{g}_{0}$, and the isovector $\bar{g}_{1}$ pion-nucleon interaction coupling constants. This results in the effective Hamiltonian
\begin{equation}
\begin{aligned}[t]
{H^{P,T}} &=                                     
\Omega \left[W_\mathrm{d}d_\mathrm{e}+W_\mathrm{s}k_\mathrm{s}\right] \\
&+ \Theta W_\mathcal{M} \left[\mathcal{M}_\mathrm{EDM} d^\mathrm{sr}_\mathrm{p} + g a_{\mathcal{M},0} \bar{g}_0+g a_{\mathcal{M},1}\bar{g}_1\right]\\
&+\mathcal{I}
\left[
W_{\mathrm{T}} k_\mathrm{T}
+W_{\mathrm{p}} k_\mathrm{p}
+W^\mathrm{m}_{\mathrm{s}}\gamma k_\mathrm{s}
+W_{\mathcal{S}}\left(ga_0 \bar{g}_0+ga_1\bar{g}_1\right)
\right.\\&\,\left.
+W^\mathrm{m}_{\mathrm{d}}\gamma d_\mathrm{e}
+W_{\mathrm{m}}\eta_{\mathrm{n}} d^\mathrm{sr}_\mathrm{n}
+W_{\mathcal{S}}R_\mathrm{vol}d^\mathrm{sr}_\mathrm{n}
\right],
\label{eq:ptodd_spinrot}
\end{aligned}
\end{equation}
where $\mathcal{I}$ is the projection of the total nuclear angular momentum $\vec{I}$ on the molecular axis, $\Theta$ is the projection of the product of $\vec{J}_\mathrm{e}$ with the second order tensor of $\vec{I}$ on the molecular axis, $W_\mathcal{M}$ is the electronic structure constant for the interaction with a NMQM, $\mathcal{M}_\mathrm{EDM}$ and $a_{\mathcal{M},0}$ $a_{\mathcal{M},1}$ are nuclear structure constants for different contributions to the NMQM, $\eta_{\mathrm{n}}=\frac{\mu_\mathrm{N}}{A}+\frac{\mu}{A-Z}$, and $\gamma$ is the nuclear gyromagnetic ratio $\gamma=\mu/I$. For $^{235}$U, $\eta_{\mathrm{n}}=0.00156\,\mu_\mathrm{N}$, and $\gamma=-0.11\,\mu_\mathrm{N}$. The $CP$ violation parameters, electronic-structure coefficients $W_i$, and nuclear-structure coefficients are defined as in Ref.~\cite{gaul:2024a}.  We provide a list of all definitions of $W_i$ in the appendix. 

We computed all electronic structure enhancement factors $W$ and compare them to selected systems which are discussed for measuring $P,T$-odd EDMs in Table \ref{tab:ptodd_cghf}. Due to an electronic ground state similar to that of \ce{PaF^{3+}} \cite{zulch:2022}, we expect a similar enhancement of symmetry violating effects, which is reflected in Table \ref{tab:ptodd_cghf}.  In \ce{UO^3+} nuclear $CP$-violation as characterized by $W_\mathcal{S}$, $W_\mathrm{m}$, $W_\mathrm{T}$, and $W_\mathrm{p}$ is largely enhanced, whereas electronic and semi-leptonic $CP$-violation sensitivity ($W_\mathrm{d}$, $W_\mathrm{s}$ and $W_\mathcal{M}$) is suppressed because of an unpaired electron of high angular momentum  with little probability density in the vicinity of the nucleus. We find a slightly larger sensitivity of electron spin dependent $P,T$-odd interactions characterized by $W_\mathrm{d}$, $W_\mathrm{s}$ and $W_\mathcal{M}$ in \ce{UO^3+} than in \ce{PaF^{3+}} but a slightly reduced sensitivity to nuclear $CP$-violation. Both, \ce{UO^3+} and \ce{PaF^{3+}}, exhibit complementary sensitivities compared to open shell systems such as \ce{ThF+} as well as to typical closed shell systems such as TlF and RaOCH$_3^+$. 

\begin{table*}[!htb]
\centering
\caption{\textit{P,T}-odd properties of $^{235}$UO$^{3+}$ obtained at the level of
2c-ZORA-cGKS-BHandH. Calculations of RaOCH$^+_3$ from Ref.~\cite{gaul:2024} were carried out at the level of 4c Dirac--Kohn--Sham (4c-DKS) using the BHandH functional. All other molecules were computed with the same methodology as employed in this work for $^{235}$UO$^{3+}$.}
\label{tab:ptodd_cghf}
\begin{ruledtabular}
\begin{tabular}{
c
c
S[table-format=-2.2,round-mode=figures,round-precision=3]
S[table-format=-3.0,round-mode=figures,round-precision=3]
S[table-format=-1.5,round-mode=figures,round-precision=3]
S[table-format=-1.3,round-mode=figures,round-precision=3]
S[table-format=-1.3,round-mode=figures,round-precision=3]
S[table-format=-2.1,round-mode=figures,round-precision=3]
S[table-format=-2.1,round-mode=figures,round-precision=3] 
}
& State
& {$W_\mathcal{S} / \frac{\mathrm{nV}}{\mathrm{fm}^3}$}
& {$W_\mathrm{m} / \frac{\mathrm{kV}}{\mathrm{cm}\mu_\mathrm{N}}$}
& {$W_\mathrm{d}/ \frac{\mathrm{GV}}{\mathrm{cm}}$}
& {$W_\mathcal{M}/ \frac{\mathrm{EV}}{c\mathrm{cm}^2}$}
& {$W_\mathrm{s}/\mathrm{peV}$}
& {$W_\mathrm{T} / \mathrm{peV}$}
& {$W_\mathrm{p} / \mathrm{feV}$}
\\
\hline
UO$^{3+}$ & $^2\Phi_{5/2}$                      & -10.4& 399 & 3.10&0.166 &20.4&-23.9&-95.4\\
\hline
TlF~\cite{gaul:2024a}           & $^1\Sigma^+$   & 6.94   & -602& 0.00665 &  0& 0.208   &  16.3 & 60.8        \\
RaOCH$_3^+$~\cite{gaul:2024}  & $^1\Sigma^+$   & -9.22  & 854  & 0.0144  &0 & 0.386   & -18.4 &-72.4\\
PaF$^{3+}$~\cite{zulch:2022,gaul:2024a}& $^2\Phi_{5/2}$     & -11.5 & 799 &2.13&  0.122      &13.4&-24.3&-97.4\\
ThF$^{+}$~\cite{gaul:2024a}& $^3\Delta_{1}$     & -7.35 & 527 &34.7&   2.51     &197&-15.2&-60.3\\
RaF~\cite{gaul:2024a} &$^2\Sigma_{1/2}$  &       -4.060160&      337.688900& -105.181248  &       -6.688770&     -585.993420 &       -7.352359 &      -29.006143 \\
\end{tabular}
\end{ruledtabular}
\end{table*}

 To estimate the impact of a possible precision experiment with \ce{UO^3+} we follow Ref.\,\cite{gaul:2024a} and perform a global analysis of the Hamiltonian (\ref{eq:ptodd_spinrot}) in a seven-dimensional $CP$-violation parameter space as recently demonstrated for AcF \cite{athanasakis-kaklamanakis:2025}.

As in previous works \cite{gaul:2024a,athanasakis-kaklamanakis:2025} we consider the following experiments: $^{129}$Xe~\cite{allmendinger:2019,sachdeva:2019}, $^{171}$Yb~\cite{zheng:2022}, $^{133}$Cs~\cite{murthy:1989}, $^{199}$Hg~\cite{graner:2016}, $^{205}$Tl~\cite{regan:2002}, $^{225}$Ra~\cite{parker:2015,bischof:2016}, $^{174}$YbF~\cite{hudson:2002,hudson:2011}, $^{180}$HfF$^{+}$~\cite{roussy:2023}, $^{205}$TlF~\cite{cho:1991}, $^{207}$PbO~\cite{eckel:2013},$^{232}$ThO~\cite{andreev:2018}, as well as the neutron experiment \cite{abel:2020}. Together with \ce{UO^3+} this results in a $13\times7$ sensitivity matrix $\bm{W}$, where the element in row $i$ and column $j$ corresponds to the Hamiltonian term in Eq.~\ref{eq:ptodd_spinrot} for atom/molecule $i$ and $P,T$-odd parameter $j$. Electronic-structure coefficients for all listed atoms and molecules were taken from Ref.~\cite{gaul:2024a} and nuclear-structure and neutron coefficients were taken from Ref.~\cite{chupp:2019}, where available. Otherwise nuclear-structure estimates reported in Ref.~\cite{gaul:2024a} were employed. For ${}^{235}$U we employ the estimate for the collective Schiff moment from Ref.\,\cite{flambaum:2020}: $g a_0=-7.8\,\mathrm{fm^3}$, $g a_1=516\,\mathrm{fm^3}$, and for the collective NMQM from Ref.\,\cite{flambaum:2022}: $g a_{\mathcal{M},0}=0.081\,\mathrm{fm^2}$, $g a_{\mathcal{M},1}=-0.405\,\mathrm{fm^2}$. 

The upper bound on $P,T$-odd constants $\vec{x}^{\mathsf{T}}=[k_{\mathrm{T}}, k_{\mathrm{p}}, k_{\mathrm{s}},\bar{g}_0,\bar{g}_1, d_{\mathrm{e}}, d_{\rm{p}}^{\rm{sr}}]$ can be determined from the experimental frequency standard uncertainties $\bm{U}=\mathrm{diag}(\delta\vec{\nu}_i^2)$ and the sensitivity matrix $\bm{W}$ by computing 
\begin{equation}
    \vec{x} > \sqrt{\left|\mathrm{diag}\left(\left[\bm{W}^\mathsf{T} \bm{U}^{-1}\bm{W}\right]^{-1}\right)\right|}\,,
\end{equation}
where $\left|\mathrm{diag}\left(\bm{A}\right)\right|$ denotes the vector of absolute diagonal elements of matrix $\bm{A}$.
Following Ref.~\cite{chupp:2015,gaul:2024a}, the restrictive power of a single experiment may be quantified by its effect on the volume $V$ enclosed by an ellipsoid described by the matrix $\bm{W}^\mathsf{T} \bm{U}^{-1}\bm{W}$. For estimating the influence of a possible experiment with \ce{UO^3+} we assume an experimental uncertainty of 1\,mHz, which is one order of magnitude larger than the uncertainty of the TlF experiments from 1991 \cite{cho:1991} and 15 times larger than the projected sensitivity of a trap experiment with a single \ce{RaOCH3+} molecule \cite{yu:2021,gaul:2024}. We summarize all results in Table~\ref{tab: limits}.

\begin{table*}[h!]
\caption{Influence of the proposed EDM experiment with $^{235}$UO$^{3+}$ on global bounds on sources of $CP$-violation assuming an experimental uncertainty of 1\,mHz. Global bounds are obtainedfrom seven-dimensional ellipsoidal coverage regions at 95\%\  confidence level, including existing experiments with $^{129}$Xe~\cite{allmendinger:2019,sachdeva:2019}, $^{171}$Yb~\cite{zheng:2022}, $^{133}$Cs~\cite{murthy:1989}, $^{199}$Hg~\cite{graner:2016}, $^{205}$Tl~\cite{regan:2002}, $^{255}$Ra~\cite{parker:2015,bischof:2016}, $^{174}$YbF~\cite{hudson:2002,hudson:2011}, $^{180}$HfF$^{+}$~\cite{roussy:2023}, $^{205}$TlF~\cite{cho:1991}, $^{207}$PbO~\cite{eckel:2013}, $^{232}$ThO~\cite{andreev:2018}, and the neutron \cite{abel:2020}. All electronic-structure parameters for these experiments were taken from Ref.~\cite{gaul:2024a} and were combined, where available, with nuclear-structure data from Ref.~\cite{chupp:2019}. In other cases, estimates of nuclear-structure parameters from Ref.~\cite{gaul:2024a} were employed. Sensitivity parameters of AcF were taken from Ref.\,\cite{athanasakis-kaklamanakis:2025}. For \ce{^{235}UO^3+} we use electronic structure calculations of the present work and for \ce{^{229}ThF+} those of Ref.\,\cite{gaul:2020a}. Nuclear structure estimates for $^{229}$Th and $^{235}$ are taken from Refs. \cite{flambaum:2020,flambaum:2022}. A conservative theoretical uncertainty of 20\%\ is assumed and the electronic-structure sensitivity factors to all sources are scaled by a factor of $0.8$ to account for a worst-case scenario.} 
\label{tab: limits}
\begin{ruledtabular}
\begin{tabular}{l
S[round-precision=0,round-mode=places,table-format=1e-2,scientific-notation=true]
S[round-precision=0,round-mode=places,table-format=1e-2,scientific-notation=true]
S[round-precision=0,round-mode=places,table-format=1e-2,scientific-notation=true]
S[round-precision=0,round-mode=places,table-format=1e-2,scientific-notation=true]
S[round-precision=0,round-mode=places,table-format=1e-2,scientific-notation=true]
}
Source
&\multicolumn{1}{c}{Current bounds}
&\multicolumn{1}{c}{With \ce{^{235}UO^3+}}
&\multicolumn{1}{c}{With \ce{^{229}ThF^+}}
&\multicolumn{1}{c}{With \ce{^{235}UO^3+} and \ce{^{227}AcF}}
&\multicolumn{1}{c}{With \ce{^{229}ThF^+} and \ce{^{227}AcF}}
\\
\hline
$V/(\mathrm{e\,cm})^2$                 &4.2e-97 & 1.6e-98 & 1.0e-98 & 2.7e-99 & 5.4e-99 \\
\hline
$d_\mathrm{e}/(e\,\mathrm{cm})$                  & 4e-29     & 4e-29 & 4e-29 & 4e-29 & 4e-29   \\
$d^\mathrm{sr}_\mathrm{p}/(e\,\mathrm{cm})$      & 1e-23     & 8e-24 & 8e-24 &  4e-24&  4e-24\\
$k_\mathrm{sps}$                  & 1e-8      & 1e-8   & 1e-8 & 1e-8& 1e-8\\
$k_\mathrm{T}$                  & 4e-6      & 4e-6  & 4e-6 & 3e-6 & 3e-6\\
$k_\mathrm{p}$                  & 9e-4      & 9e-4  & 9e-4 & 9e-4 & 9e-4\\
$\bar{g}_\mathrm{0}$            & 6e-10      & 5e-10 & 5e-10 & 3e-10& 3e-10\\
$\bar{g}_\mathrm{1}$            & 2e-10      & 1e-10 & 1e-10 & 6e-11& 6e-11\\
\end{tabular}       
\end{ruledtabular}
\end{table*}
 
It should be noted that a precision better than 1\,mHz in the proposed \ce{UO^3+} experiment would not further improve global constraints on individual parameters, as these are limited by the experiments with the largest uncertainties (see also Ref.~\cite{degenkolb:2024} for a discussion). This demonstrates the urgent need for more experiments, which are sensitive to nuclear $CP$-violation, as for example \ce{UO^3+}. 
For comparison with AcF in Ref.\,\cite{athanasakis-kaklamanakis:2025} it must be noted that in Ref.\,\cite{athanasakis-kaklamanakis:2025} the neutron experiment was not included in the analysis. Excluding the neutron experiment would increase the impact of a \ce{UO^3+} experiment. In order to demonstrate the importance of complementarity over individual sensitivity we compare to a possible experiment with $^{229}$ThF$^+$ assuming the same uncertainty. Both experiments lead to similar bounds. As we employ the estimates of Refs. \cite{flambaum:2020,flambaum:2022} no complementary sensitivity between $\bar{g}_0$ and $\bar{g}_1$ can be achieved. Inclusion of more sophisticated nuclear structure calculations can change relative sizes of enhancement factors, as for example computed recently for AcF in Ref.\,\cite{athanasakis-kaklamanakis:2025}. This has a large impact on the projected bounds. The overall volume reduction and improvement of global bounds by AcF is similar or less pronounced than for  \ce{^{235}UO^3+} or \ce{^{229}ThF^+}. It is likely, that more sophisticated nuclear structure calculations will reveal complementary sensitivities between \ce{^{235}UO^3+} and \ce{^{229}ThF^+} in the nuclear sector. Moreover, it is important to recognize that an experiment with \ce{UO^3+}, using the estimate of Ref.\,\cite{flambaum:2020} for the nuclear Schiff moment, would be more sensitive to $\bar{g}_1$ than the Hg experiment when an experimental uncertainty of 0.1\,mHz can be realized.

\subsubsection{Search for temporal variations of fundamental constants}

We expect a  comparatively simple but congested level-structure in \ce{UO^3+} similar to \ce{PaF^3+} \cite{zulch:2022,zulch:2025}, which may offer opportunities to search for variations of fundamental constants  due to accidentally quasi-degenerate vibronic levels of different symmetry in the electronic degrees of freedom \cite{flambaum:2007,zulch:2025}.

\section{Conclusion and Outlook}

In this work, we combined HFLA with complementary MR-TOF mass-spectrometric techniques to investigate the formation and stability of uranium-containing molecular ions. MR-TOF measurements established the isotopic enrichment and high purity of the uranium foils , while the corresponding TOF spectra showed no evidence for low-mass contaminants. Laser ablation of the uranium foils yielded in mono and diuranium atomic and molecular cations. 

In contrast, HFLA enables the direct production of multiply charged uranium monoxide ions \ce{UO^z+} up to charge state $z=4$. This includes the open shell ion \ce{UO^3+}, which is isoelectronic to \ce{RaF}. With relativistic DFT calculations we demonstrated favorable electronic structure properties and large enhancements of $P,T$-violation, which render \ce{UO^3+} an interesting candidate for precision searches for $CP$-violation beyond the Standard Model. Combined with intrinsically octupole deformed odd uranium isotopes, which are easier to access than Pa or Th isotopes, \ce{UO^3+} positions as a uniquely powerful probe of symmetry violating as well as conserving nuclear moments. 

In addition, we report the observation of the \ce{UO^4+} ion, which pushes actinide molecular chemistry to the edge of chemical stability. Relativistic DFT calculations confirmed the metastability of the produced ions, quantified their charge distributions and provided insights into the electronic configurations that lead to chemical bonds under extreme conditions.

The presented results elevate uranium monoxide ions from a previously unexplored species to a versatile platform at the intersection of chemistry, spectroscopy, and precision measurements. The combination of experimental accessibility, theoretical predictability, and exceptional sensitivity to new physics makes \ce{UO^3+} and \ce{UO^4+} compelling systems for precision spectroscopy.
 From an experimental perspective, contaminant loading in future trap-based experiments can be strongly suppressed by time-gated ion injection. Since the relevant species are separated in time of flight on the microsecond scale, fast switching of a trap entrance or endcap electrodes can be used to preferentially load a selected mass-to-charge window. For example, U$^{4+}$, UO$^{4+}$, and UO$^{3+}$ were detected at 44.2 $\mu$s, 45.5 $\mu$s, and 51.7 $\mu$s. Nevertheless, internal-state preparation of the selected molecular ions remains an open experimental challenge and will require dedicated cooling and optical-pumping schemes.
The present approach is transportable and can provide multiply charged molecular ions at the edge of chemical stability for next-generation studies of $CP$ violation, nuclear structure, and possible variations of fundamental constants. The demonstrated method is not restricted to $^{238}$U. In principle, all sufficiently long-lived uranium isotopes can be employed as targets, enabling systematic studies of isotope-dependent nuclear moments and symmetry-violating effects. This opens the door to exploring how nuclear deformation, octupole collectivity, and higher-order nuclear moments manifest along the uranium isotopic chain in a molecular environment.
More broadly, the demonstrated ability of HFLA to reproducibly generate such exotic molecules establishes it as a universal tool for producing and probing multiply charged actinide species. Extending this approach to neighboring actinides, such as neptunium monoxide (\ce{NpO^4+}) and plutonium monoxide (\ce{PuO^4+}), promises to open further opportunities at the frontier of nuclear structure, molecular spectroscopy, and searches for physics beyond the Standard Model.

\section{Acknowledgements}
We express our gratitude to Bettina Lommel and Birgit Kindler from GSI for purchasing and providing the depleted uranium foil.

This work has been supported by the Cluster of Excellence “Precision Physics, Fundamental Interactions, and Structure of Matter$^+$” (PRISMA$^+$ + ExNet-020). Funded by the German Research Foundation (DFG) under TACTICa (Project Nr. 495729045). K.G. thanks the Fonds der Chemischen Industrie (FCI) for generous funding through a Liebig fellowship. We gratefully acknowledge computing time at the supercomputer MOGON 2 at Johannes Gutenberg University Mainz (hpc.uni-mainz.de), which is a member of the AHRP (Alliance for High Performance Computing in Rhineland Palatinate,  www.ahrp.info) and the Gauss Alliance e.V. P.F. and L.S. acknowledge funding from the German Ministry for Education and Research (BMBF) under Grant No. 05P18HGCIA. Fruitful discussions with Simon Benedikt Diewald are acknowledged.

\section{Author contributions}
J.S.: conceptualization, formal analysis, investigation, writing - original draft. 
P.F.: formal analysis, investigation, writing - review \& editing.
K.G.: resources (theoretical), funding acquisition, formal analysis, investigation (theoretical), writing- original draft. 
L.M.A.: formal analysis, investigation.
F.K.: resources, writing - review \& editing. 
D.K.: resources, review \& editing. 
D.R.: resources, writing - review \& editing. 
F.S.-K.: funding acquisition, supervision, writing - review \& editing.
L.S.: resources, funding acquisition, supervision, writing- review \& editing. 
J.V.: formal analysis, investigation.
C.E.D.: resources, funding acquisition, supervision, writing - review \& editing.

\bibliography{References}
\FloatBarrier

\end{document}